\documentclass[10pt, conference, letterpaper]{IEEEtran}

\usepackage{booktabs} 
\usepackage{enumerate}
\usepackage{graphicx}
\usepackage[ruled, vlined, linesnumbered]{algorithm2e}
\usepackage{comment}
\usepackage{amsmath}
\usepackage{amsthm}
\usepackage[outdir=./]{epstopdf}
\usepackage{subfigure}
\usepackage{float}
\usepackage{makecell}
\usepackage{cite}
\usepackage[hyphens]{url}
\usepackage[hidelinks]{hyperref}
\begin{document}
\title{Traffic Analytics Development Kits (TADK): \\ Enable Real-Time AI Inference in Networking Apps}

\author{
  \IEEEauthorblockN{Kun Qiu\IEEEauthorrefmark{1}, Harry Chang\IEEEauthorrefmark{1}, Ying Wang\IEEEauthorrefmark{1}, Xiahui
    Yu\IEEEauthorrefmark{1}, Wenjun Zhu\IEEEauthorrefmark{1}, Yingqi Liu\IEEEauthorrefmark{1}, Jianwei Ma\IEEEauthorrefmark{1}, Weigang Li\IEEEauthorrefmark{1}}
  \IEEEauthorblockN{Xiaobo Liu\IEEEauthorrefmark{2}, Shuo Dai\IEEEauthorrefmark{2}}
 \IEEEauthorblockA{\IEEEauthorrefmark{1}Intel Corporation, \IEEEauthorrefmark{2}ZTE Corporation}

 \{kun.qiu,harry.chang,ying.wang,xiahui.yu,wenjun.zhu,yingqi.liu,jianwei.ma,weigang.li\}@intel.com, \\\{liu.xiaobo2,dai.shuo\}@zte.com.cn}

\maketitle

\begin{abstract}
  Sophisticated traffic analytics, such as the encrypted traffic analytics and unknown malware detection, emphasizes the need for advanced methods to analyze the network traffic. Traditional methods of using fixed patterns, signature matching, and rules to detect known patterns in network traffic are being replaced with AI (Artificial Intelligence) driven algorithms. However, the absence of a high-performance AI networking-specific framework makes deploying real-time AI-based processing within networking workloads impossible.
  In this paper, we describe the design of Traffic Analytics Development Kits (TADK), an industry-standard framework specific for AI-based networking workloads processing. TADK can provide real-time AI-based networking workload processing in networking equipment from the data center out to the edge without the need for specialized hardware (e.g., GPUs, Neural Processing Unit, and so on). We have deployed TADK in commodity WAF and 5G UPF, and the evaluation result shows that TADK can achieve a throughput up to $35.3$\textit{Gbps} per core on traffic feature extraction, $6.5$\textit{Gbps} per core on traffic classification, and can decrease SQLi/XSS detection down to $4.5 \mu s$ per request with higher accuracy than fixed pattern solution.
\end{abstract}

\section{Introduction}
Silicon and software technology advancements targeting AI inference have lowered the barrier (compute cost and R\&D effort) to unleash the creativity and innovation of the network application developers on the use of AI-advanced techniques within their commercial solutions.
 Reports and analysis are projecting the use of AI in Enterprise SD-WAN deployments to increase from $5\%$ in 2021 to $40\%$ in 2025~\cite{SD-WAN}.

Industry practices are introducing AI techniques using artificial intelligence (AI) and machine learning (ML) models across network analytics approach. Here are some examples of use cases: (1) \textbf{Traffic analytics:} Used to analyze encrypted network traffic, to identify anomalies within networks~\cite{nguyen2008survey}; (2) \textbf{Malware Detection:} Detecting malicious traffic such as SQL injection or Cross-Site Script~\cite{dainotti2012issues}; (3) \textbf{User Behavior analytics:} Detecting relationships, identifying anomalies, and conducting empirical assessments of
security~\cite{megyesi2015user,molnar2013multi,vassio2016detecting}.

In order to support real-world workloads, an industry-standard framework for real-time AI traffic analytics has to meet the requirements for performance, accuracy, and scalability. Based on previous research and discussion with our customers and partners, we have identified several mutually challenging as follows:

\begin{itemize}
  \item \textbf{High Throughput:} up to $3$\textit{Gbps} per core (rule-based level) for AI-based traffic classification~\cite{wang2019hyperscan,hybench}
  \item \textbf{Low Latency:} $5\sim10 \mu s$ per request for malicious traffic detection~\cite{libin}
  \item \textbf{High Accuracy:} $\ge 95\%$ accuracy
  \item \textbf{Easy Deployment:} deploying without the need for specialized hardware (e.g., GPU, NPU, FPGA)
  \item \textbf{Easy Development:} module-based development like DPDK~\cite{DPDK} and VPP~\cite{VPP}
\end{itemize}

To address above challenges, we have designed Traffic Analytics Development Kits (TADK), an industry-standard framework specific for AI-based networking workloads processing. TADK can provide real-time AI-based networking workload processing in networking equipment from the data center out to the edge without the need for specialized hardware~\cite{tadkincpe}. Briefly speaking, TADK brings several advantages to AI-based networking processing:

\begin{enumerate}
  \item \textbf{High Performance:} TADK provides highly-optimized library for real-time AI-based traffic analytics. We design several novel algorithms to increase the performance. From our benchmarking results, traffic classification can achieve up to $6.5$\textit{Gbps} per core, which can fully support real-time classification in most cases. Meanwhile, the overall pipeline of SQLi/XSS detection can achieve up to $4.5\sim6.1 \mu s$ per HTTP request, which is $2$x faster than the existing rule-based solution. Also, the accuracy of traffic classification and SQLi/XSS detection is $\ge 95\%$
  in most cases~\cite{barut2021multi,barut2020netml,barut2020tls,barut2020machine}.
  \item \textbf{Easy Deployment:} The application developed with TADK does not rely on any specialized hardware. TADK fully utilizes modern CPU features such as \textbf{AVX512} to accelerate AI performance.
  \item \textbf{Easy Development:} TADK offers a module-based development environment. Developers can implement their own AI-based traffic analytics application by combining TADK's modules like building block bricks~\cite{tadkinwaf}.
\end{enumerate}

\begin{figure*}[!htp]
  \centering
  \includegraphics[width=6.5in]{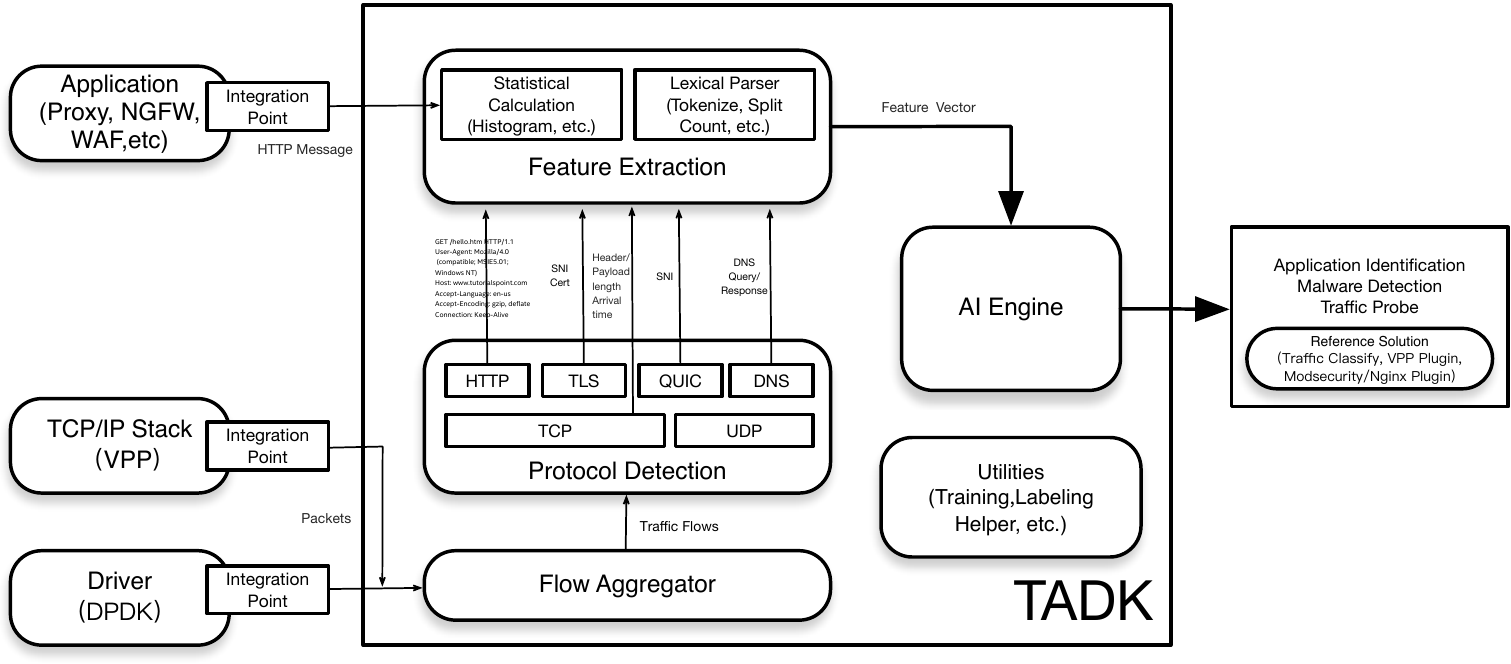}
  \caption{The overall design of TADK.}
  \label{fig:overall}
\end{figure*}

The rest of the paper is organized as follows: we first give the background and related work of AI-based traffic analytics in Section~\ref{sec:bak}. In Section~\ref{sec:design}, we will give the overview design of TADK. Then, we
will give some detail of our highly optimized feature extraction algorithms in Section~\ref{sec:feature}, and we will evaluate TADK in two scenarios: traffic classification and SQLi/XSS detection in Section~\ref{sec:eva}. We conclude in Section~\ref{sec:con}.

\section{Background and Related Work}

\label{sec:bak}

\subsection{Data Collection}
A systematic survey has concluded a general pipeline of AI-based traffic analytics~\cite{pacheco2018towards}. The first step is data collection~\cite{megyesi2015user,molnar2013multi}.
The AI-based solution needs historical data as the input source to train the model.
However, capturing and labeling enough traffics is hard to conduct, mainly due to accuracy and privacy concerns. 
It is reported that $60\%$ of research is using public non-encrypted traffic~\cite{pacheco2018towards} and using DPI tools to label traffic. In order to cover this issue, TADK provides a labeling helper that can help users to label non-encrypted and encrypted traffic with only one click.
\subsection{Feature Extraction}
The next step is called feature extraction.
The most common trend uses statistical-based features (e.g., inter-arrival time and packet size with the minimum, maximum and average metrics) since they can be used both on non-encrypted and encrypted traffic analytics~\cite{davis2011data,davis2016automated,marnerides2014traffic}.  
However, some open-source feature extraction libraries~\cite{Joy} whose performance is as not good as TADK's library. Meanwhile, TADK can extract not only statistical features but also lexical features from encrypted traffic. It is proved that the combination of statistical and lexical features can significantly increase the accuracy. The flow extraction library of TADK has been utilized in AI traffic analytics ~\cite{barut2021multi,barut2020netml,barut2020tls,barut2020machine}.
TADK provides a tokenizer that is remarkably faster than existing solutions to extract lexical features.
\subsection{AI Inference}
At the last step, an AI model or an ensemble of models are needed for gathering analytics results~\cite{peng2016effectiveness,alshammari2015identification}. Both supervised and unsupervised methods are widely deployed in traffic analytics. Labeled datasets are used to train a supervised model such as SVM, decision tree, and random forest. Unsupervised models such K-Means are utilized in anomalous traffic detection. Meanwhile, most solutions use the unsupervised model to cluster encrypted traffics since labeling encrypted
traffics~\cite{goseva2014characterization,belavagi2016performance,lalitha2016traffic} is difficult. In TADK, we provide an optimized random forest model for AI inference. We have compared a variety of models and found the random forest is well-balanced between accuracy and latency in traffic analytics workload.

\begin{figure*}[!htp]
  \centering
  \includegraphics[width=7in]{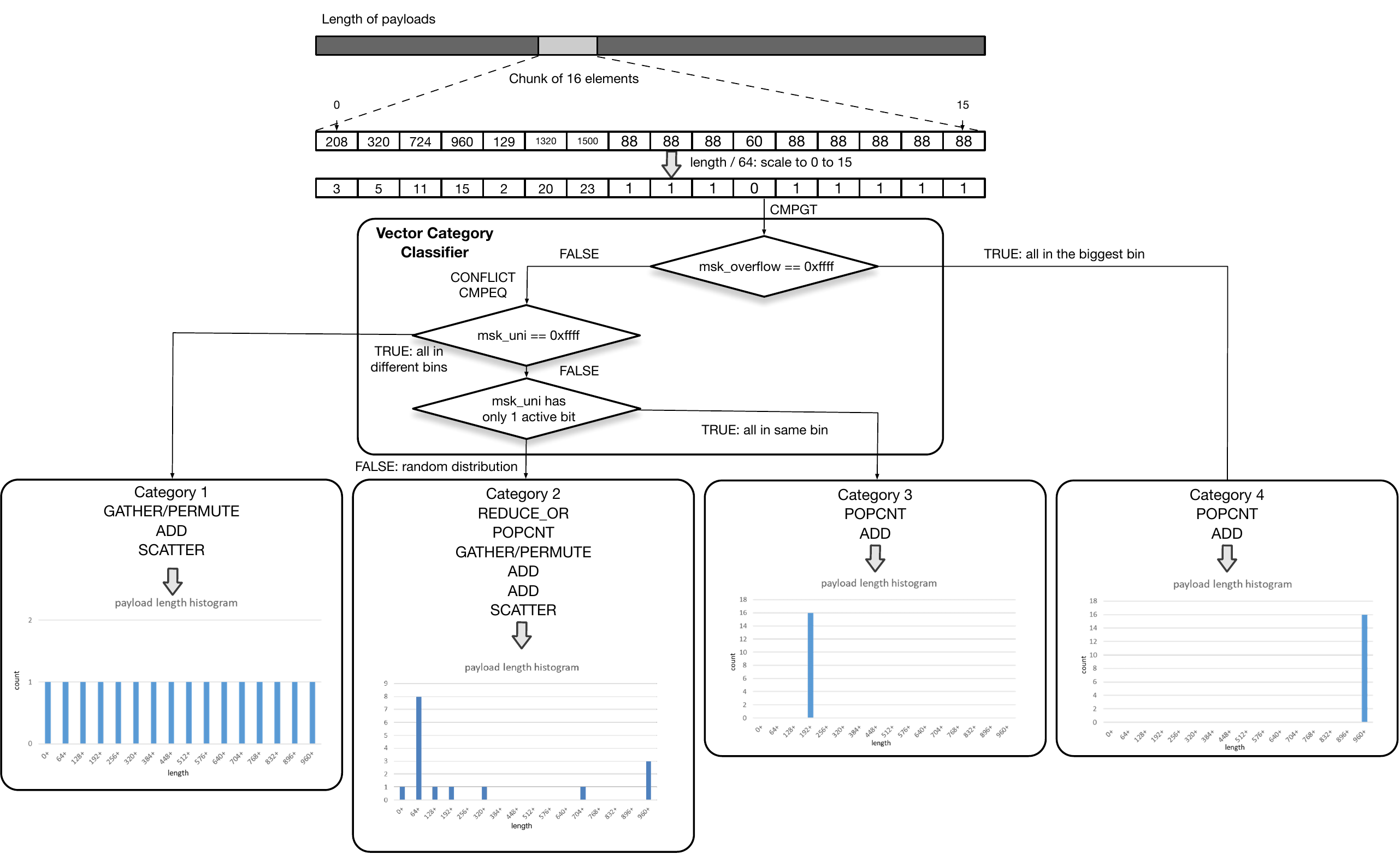}
  \caption{The workflow of Advanced Vector Calculation with Vector Category Classifier}
  \label{fig:avc}
\end{figure*}

\section{The Overall Design of TADK}
\label{sec:design}
\subsection{Core Libraries}
TADK is composed of a series of core libraries, which are corresponding to feature extraction and AI inference steps we mentioned before. We show each component in Fig.~\ref{fig:overall}.
Flow aggregator is used to aggregate traffics from packets (e.g., real-time packets or packet traces from PCAP files) by 5-tuples. Protocol detection is used to identify protocols such as TCP, TLS, QUIC, and so on.
Feature extraction, which is the competitiveness of TADK, has been well-designed to support real-time AI-based traffic analytics. We will describe some core algorithms in Section~\ref{sec:feature}.
AI engine is a wrapper of a high-performance random forest, which is based on Intel oneDAL~\cite{onedal}. Our AI engine supports both training and inferencing, including automatic feature reduction.
\subsection{Utilities}
TADK brings some useful utilities for training. The data cleaner and labeling helper provide an one-click solution for traffic labeling. The user only needs to capture one or several packet traces (e.g., PCAP files) as input of the labeling helper, and the helper will cluster these packet traces into several clusters. Each cluster will have a labeling tip. The only work for the user is to label each cluster with tips and use labeled traffic to train a model.
\subsection{Reference Solutions}
TADK provides some samples to show the reference usage of TADK core libraries.  The traffic classification sample can monitor network traffic and identify different applications in encrypted traffic.
Either packet traces (PCAP files) or real-time traffic can be the input of the traffic classification sample. The SQL injection (SQLi)/Cross-Site Script (XSS) detection sample can detect whether the payload of HTTP traffic contains malicious code. TADK also provides a VPP plugin for the traffic classification sample and ModSecurity plugin for the SQLi/XSS detection sample. With these plugins, users can directly integrate AI-based solutions into their existing pipeline without any modification. We give the integration points in Fig.~\ref{fig:overall}.

\section{Feature Extraction}
\label{sec:feature}

\subsection{SIMD-based Histogram}
Histogram, such as the distribution characteristic of TCP packet header length, payload length, and arriving time intervals, etc., are mostly used as statistical features. Thus, designing an efficient histogram algorithm is a critical issue. We take histogram calculation of TCP packet payload length as an example to illustrate the detailed implementation. A buffer of lengths of packets as a shown example in Fig.~\ref{fig:avc} used to store the payload length of each packet in a network flow (for simplicity, 16 packets are considered here). The purpose of the histogram is to count the number of each element in the buffer belonging to a specific bin.
\subsubsection{Existing Solution}
Scalar Calculation (SC) is a widely utilized method. It has been implemented in most feature extraction libraries. 
SC is a loop-based method, which means they use huge amounts of loop and branch (it has to process and count each element one by one) for the histogram. In order to cover the disadvantage, a loop-free design such as a SIMD-based algorithm has been proposed. 

\subsubsection{Advanced Vector Calculation}
We propose a SIMD-based algorithm called Advanced Vector Calculation (AVC).
As shown in Fig.~\ref{fig:avc}, we separate the input traffic into $4$ categories:
\begin{enumerate}
  \item \textbf{Category 1:} All elements are in different bins.
  \item \textbf{Category 2:} All elements are random distribution.
  \item \textbf{Category 3:} All elements are in one bin (except the biggest bin).
  \item \textbf{Category 4:} All elements are in the biggest bin.
\end{enumerate}

\begin{figure*}[!htp]
  \centering
  \includegraphics[width=5in]{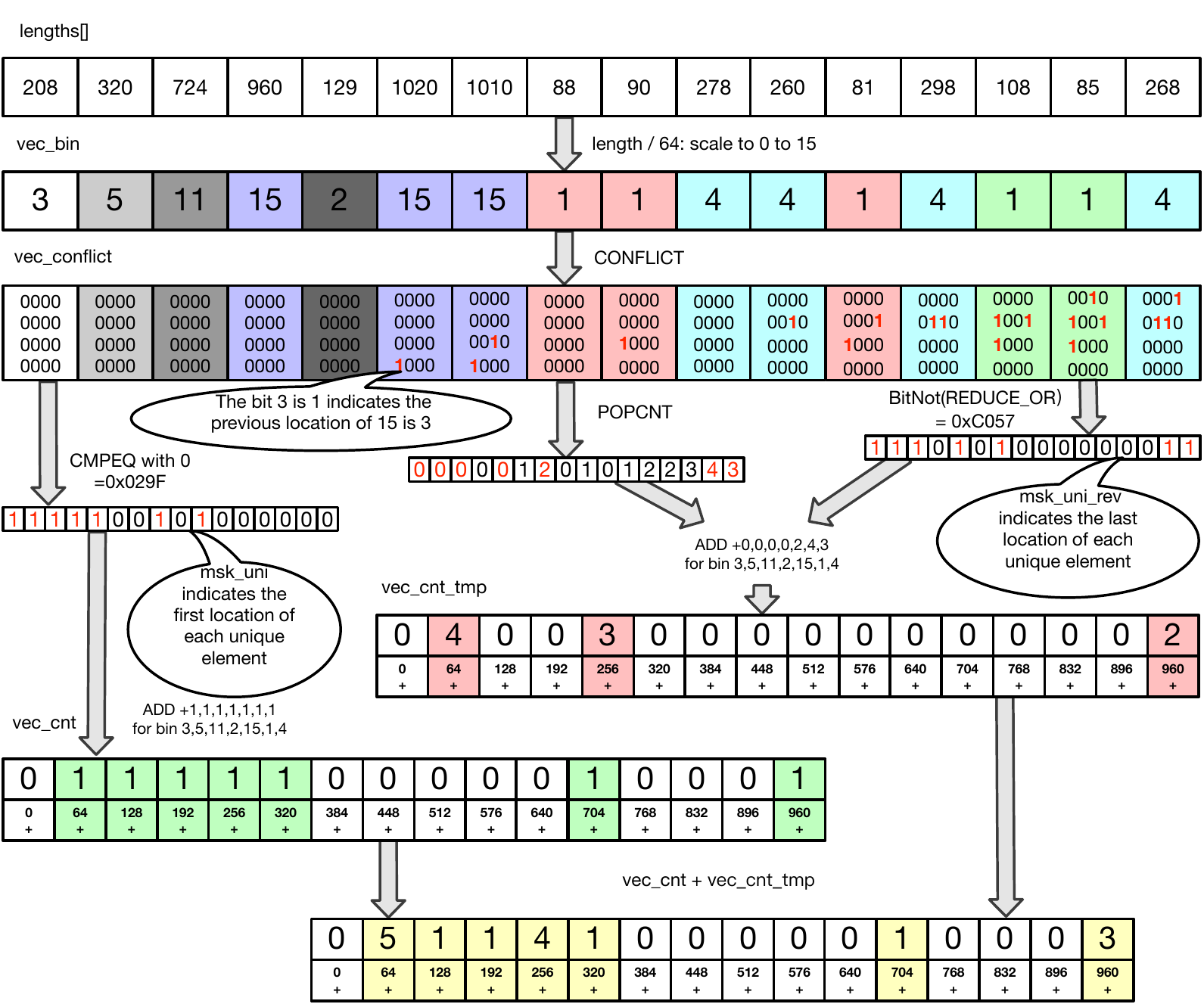}
  \caption{Calculating histogram of category 2}
  \label{fig:category2}
\end{figure*}

Since each category needs a different algorithm to
calculate, we also propose a Vector Category Classifier (VCC) to identify the category of input data. In order to prevent VCC to be an overhead of histogram calculation, we use only up to $3$ instructions to identify the category, which is also shown in Fig.~\ref{fig:avc}. We give define SIMD intrinsics in TABLE~\ref{tab:def}.
\begin{table}[!htp]\footnotesize
\caption{Definition of SIMD Intrinsics~\cite{intelins}}
\centering
\begin{tabular}{|c|l|}
\hline
\rule[-6pt]{0mm}{18pt}Notation & Description\\
\hline
\rule[-6pt]{0mm}{18pt} \textsc{CMPGT}($\vec{a}$, $\vec{b}$) & Compare $\vec{a}$ with $\vec{b}$ for greater-than\\
\hline
\rule[-6pt]{0mm}{18pt} \textsc{CONFLICT}($\vec{a}$) & Test each element of $\vec{a}$ for equality\\
\hline
\rule[-6pt]{0mm}{18pt} \textsc{REDUCE\_OR($\vec{a}$)} & Reduce each element in $\vec{a}$ by bitwise \textsc{OR}\\
\hline
\rule[-6pt]{0mm}{18pt} \textsc{CMPEQ}($\vec{a}$, $\vec{b}$) & Compare $\vec{a}$ with $\vec{b}$ for equal\\
\hline
\rule[-6pt]{0mm}{18pt} \textsc{POPCNT}($\vec{a}$) & \makecell[l]{Count the number of \\logical 1 bits in each element in a}\\
\hline
\rule[-6pt]{0mm}{18pt} \textsc{ADD}($\vec{a}$, $\vec{b}$) & Add $\vec{a}$ with $\vec{b}$\\
\hline
\rule[-6pt]{0mm}{18pt} \makecell{\textsc{GATHER}($\vec{a}$, $b$)\\\textsc{PERMUTE}($\vec{a}$, $b$)} & Load $b$ to $\vec{a}$ with a specific order\\
\hline
\rule[-6pt]{0mm}{18pt} \textsc{SCATTER}($\vec{a}$, $b$) & Store $\vec{a}$ to $b$ with a specific order\\
\hline
\end{tabular}
\label{tab:def}
\end{table}

Briefly speaking, we first use a \textsc{CMPGT} to identify whether
each element are larger than the biggest bin. If all elements is larger than the biggest bin, it is category 4. Then, we use \textsc{CONFLICT} to compute
\textit{vec\_conflict} and \textit{msk\_uni} for checking whether each element is unique. If all elements are unique 
, it is category 1. At last, we can simply check whether the \textit{msk\_uni} with only one active bit. 
If there is only one bit in the \textit{msk\_uni}, it is category 3, otherwise, it is category 2.


\begin{algorithm}
  \caption{Advanced Vector Calculation (AVC)}
  \KwIn{\textit{length} the buffer of payload length}
  \KwOut{\textit{hist} the histogram of payload length}
  \DontPrintSemicolon
  Reset \textit{hist}\;
  Load \textit{length} to \textit{vec\_len}\;
  \textit{vec\_bin} $\leftarrow$ $\frac{\textit{vec\_len}}{64}$\;
  \textit{msk\_overflow} $\leftarrow$ \textsc{CMPGE}(\textit{vec\_bin}, $\vec{15}$)\;
  \If {\textit{msk\_overflow} = \textit{0xffff}}
  {
      \textbf{Category 4: all in the biggest bin}\;
      \textit{hist}[$15$] $\leftarrow$ \textit{hist}[$15$] + $16$\;
      return \textit{hist}\;
  }
  Remove elements that is larger than $15$ in \textit{vec\_bin}\;
  \textit{vec\_conflict} $\leftarrow$\textsc{CONFLICT}(\textit{vec\_bin})\;
  Use \textsc{CMPEQ} to convert \textit{vec\_conflict} to \textit{msk\_uni}\;
  \If {\textit{msk\_uni} = \textit{0xffff}}
  {
      \textbf{Category 1: all in different bins}\;
      \textsc{GATHER}(\textit{vec\_cnt}, \textit{hist})\;
      \textit{vec\_cnt\_added} $\leftarrow$ \textsc{ADD}(\textit{vec\_cnt}, $\vec{1}$)\;
      \textsc{SCATTER}(\textit{hist},\textit{vec\_cnt\_added})\;
  }
  \ElseIf {\textit{msk\_uni} \textit{BitAnd} (\textit{msk\_uni} - 1) = $0$}
  {
      \textbf{Category 3: all in the same bin}\;
      \textit{hist}[\textit{vec\_bin}[$0$]] $\leftarrow$ \textit{hist}[\textit{vec\_bin}[$0$]] + $16$\;
  }
  \Else
  {
      \textbf{Category 2: random distribution}\;

      \textit{msk\_uni\_rev} $\leftarrow$ \textit{BitNot}     \textsc{REDUCE\_OR}(\textit{vec\_conflict})\;

      \textit{vec\_popcnt} $\leftarrow$ \textsc{POPCNT}(\textit{vec\_conflict})\;
      \textsc{GATHER}(\textit{vec\_cnt}, \textit{hist})\;
      \textit{vec\_cnt\_tmp} $\leftarrow$
      \textsc{ADD}(\textit{vec\_cnt}, $\vec{1}$)\;
      \textit{vec\_cnt\_added} $\leftarrow$ \textsc{ADD}(\textit{vec\_cnt\_tmp}, \textit{vec\_popcnt})\;
      \textsc{SCATTER}(\textit{hist},\textit{vec\_cnt\_added})\;
  }

\label{algo:1}
\end{algorithm}

Although it is easy to calculate the histogram in categories 1, 3, and 4 with up to $3$ instructions, designing an algorithm for category 2 is the most challenging work. Thus, we propose a novel algorithm to calculate the histogram in category 2. We also give an example in Fig.~\ref{fig:category2}. Algorithm~1 shows the pseudo-code of AVC and VCC. We evaluate our proposed AVC for histogram calculation. AVC can achieve up to $11.73$x, $4.38$x, $1.33$x and $1.47$x faster than the existing solution in categories 1,2,3,4 respectively.

\subsection{DFA-based Tokenization}
\begin{figure}[!htp]
  \centering
  \includegraphics[width=3.4in]{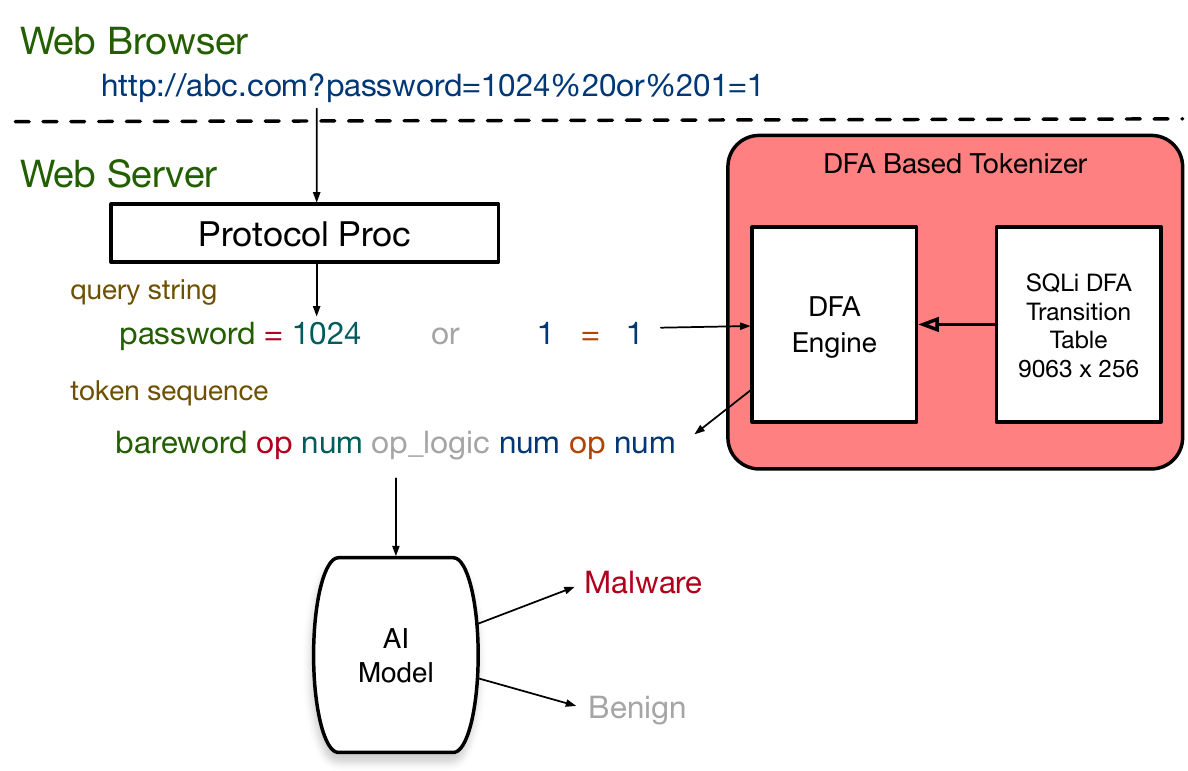}
  \caption{An example of SQLi detection with DFA-based tokenizer}
  \label{fig:SQLi}
\end{figure}

Most AI-based traffic analytics (e.g, Next Generation Web Application Firewalls) needs tokenization to convert lexical features (string-based information) into vectors as the input of the AI-model. For lexical features, most tokenizers (e.g., OpenNMT) are branch-based, which means they use huge amounts of IF-ELSE for tokenizing. A branch-based solution is easy to implement, but it is unfriendly to the CPU’s pipeline, and it may increase the number of cache misses. Thus, TADK uses a DFA-based tokenizer and provides a generator that can convert an easy-to-code profile into a specific DFA. We give an example of SQLi detection with a DFA-based tokenizer in Fig.~\ref{fig:SQLi}. We also propose a training video~\cite{sqlivideo} to describe how the tokenizer works.

\subsubsection{Generator}
In order to support multiple language/file formats, we propose a generator that can generate DFA from user-defined profiles. We defines a DFA profile language which can be easily maintained by our customers, and easy to extend to add new tokens for emerging threats, and to support more use cases. The generator also includes a DFA compiler to compile the user-defined profile into its corresponding DFA transition table. The DFA transition table is directly used by the Tokenizer.

\subsubsection{Tokenizer}
\begin{algorithm}
  \caption{DFA engine}
  \KwIn{\textit{T} the transition table\\
        \textit{V} the input string\\
        \textit{A} the state table}
  \KwOut{\textit{R} the accept state}
  \DontPrintSemicolon
  $S \leftarrow $ inital state\;
  \For {$i \in$ the numbers of character \textit{C} divide by \textit{V}}
  {
       $S \leftarrow T[S][C]$\;
       \If {$S$ is accept state}
       {
            output $A[S]$\;
       }
  }
\label{algo:2}
\end{algorithm}

The DFA transition table describes the transition behavior under every state and input character. Algorithm~2 shows how the DFA engine works. The engine does simple transitions in the main loop which makes it very fast.

\section{Evaluation}
\label{sec:eva}

\subsection{Environment}
We implement TADK using GCC 7.5. Since TADK has been deployed in several scenarios, such as WAF or 5G User Plane Function (UPF), we have different CPU and RAM environment. The 5G UPF uses ZTE 5300G4X, which is based on Intel Xeon Gold 6330N CPU (Icelake) with 512GB DDR4 RAM. Other evaluation is based on Xeon Gold 6148 CPU (Skylake) and Intel Xeon Platinum 8358 CPU (Icelake) with 32G DDR4 RAM.
We integrate our reference traffic classification sample into ZTE 5G UPF to test its throughput and accuracy. We use IXIA as a traffic generator to generate traffic to test the maximum throughput with zero packet loss.

\subsection{Data}
Since we choose random forest as our AI inference model, we evaluated the accuracy of random forest in both traffic classification and malware detection. In traffic classification, we have collected top applications in China (BAIDU, TMALL, BILIBILI, TENCENT, TOUTIAO, KUAISHOU, QQ, HUOSHAN, QQNEWS, YOUKU, WECHAT)
from the real-world, for both training and inferencing. In malware detection, we use \textsc{SQLMAP}~\cite{sqlmap} for SQLi and \textsc{XSStrike}~\cite{xss} for XSS to gather data for both training and inferencing. We also choose some public data for inferencing. 

\subsection{Traffic Classification}
\subsubsection{Accuracy}
\begin{figure}[!htp]
  \centering
  \includegraphics[width=3.4in]{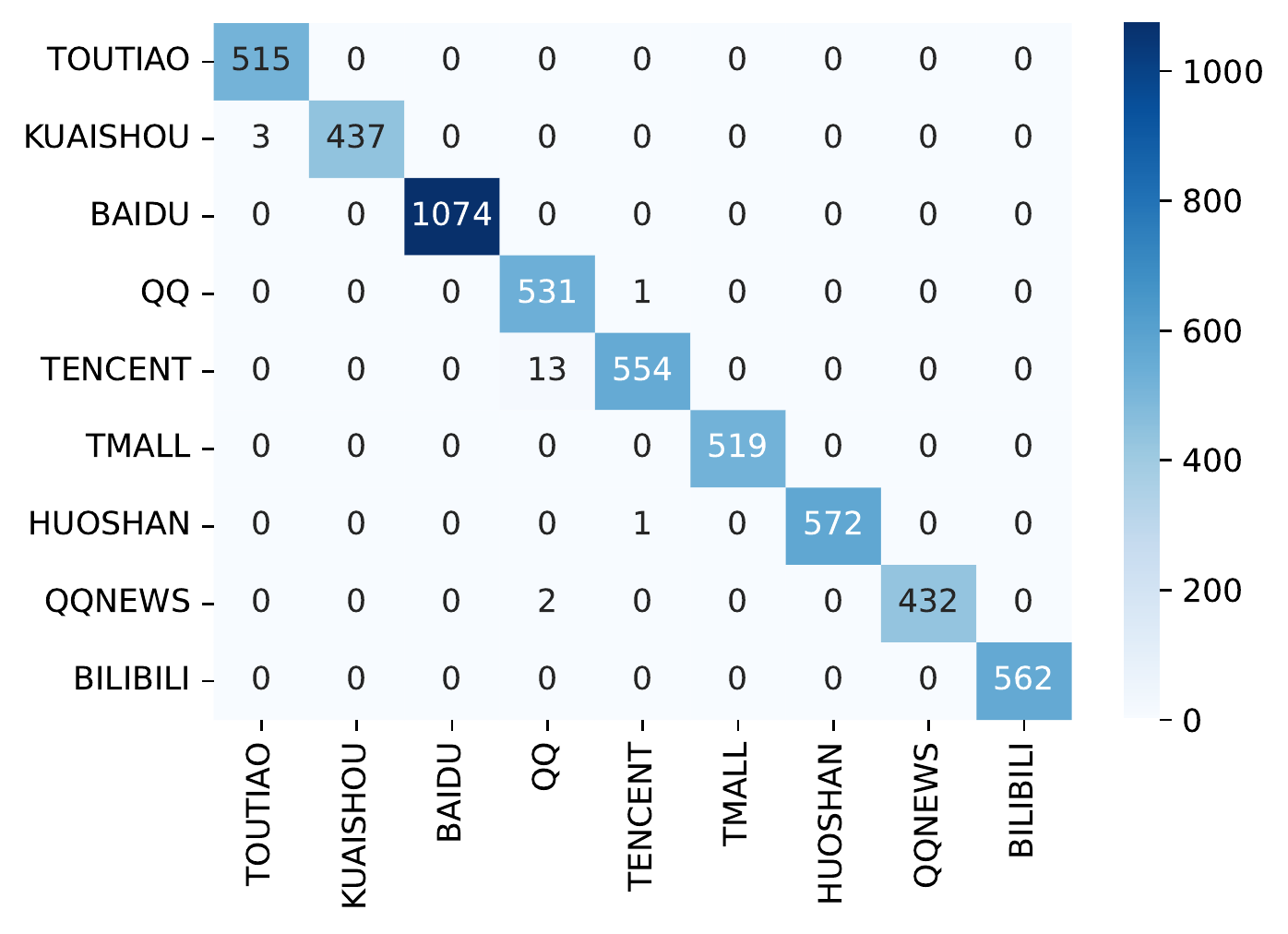}
  \caption{The confusion matrix of $9$ applications}
  \label{fig:conf}
\end{figure}

We give the confusion matrix of the model that can classify $9$ applications in Fig.~\ref{fig:conf}.
All precision and recalls are larger than $90\%$, and the average precision, recall, and f1-score are $0.936, 0.926, 0.918$. From the evaluation result we can see that the accuracy for traffic classification is
sufficient for most scenarios. We also train a model to classify WeChat image transfer traffic and WeChat video transfer traffic, which are UDP traffic.  We prepare $70$ image transfer flows and $100$ video transfer flows to train, and we give the accuracy detail in TABLE~\ref{tab:wc}. The average precision, recall, and f1-score are $0.883, 0.884, 0.883$.

\begin{table}[!htp]\footnotesize
\caption{Classify WECHAT Video and Image Transfer}
\centering
\begin{tabular}{|c|c|c|c|c|}
\hline
 Class & Precision & Recall & F1-score & Flows\\
\hline
 WECHAT Video & $0.903$ & $0.875$ & $0.889$ & $32$\\
\hline
WECHAT Image & $0.862$ & $0.893$ & $0.877$ & $28$\\
\hline
\end{tabular}
\label{tab:wc}
\end{table}

\subsubsection{Performance}
We test our latency with the model that can classify $2$ applications (train and test by WECHAT with $1524$ flows and YOUKU with $1551$ flows). From Table~\ref{tab:etala} we can see that our latency can achieve $10.7\mu s$ per flow, which is sufficient for most scenarios. Moreover, we also test the latency of feature extraction for DNS, HTTP and TLS in Table~\ref{tab:etala}. The average packets of DNS, HTTP, and TLS are $2$, $8$ and $13$. With the \textsc{POPCNT} instruction and the new architecture, the latency has been reduced significantly. The reason why TLS has lower latency than HTTP is TLS has less lexical features to extract.

\begin{table}[!htp]\footnotesize
\caption{Latency per flow}
\centering
\begin{tabular}{|c|c|c|c|c|c|}
\hline
 & \multicolumn{2}{c|}{Traffic Classification} & \multicolumn{3}{c|}{Feature Extraction}\\
\hline
 Architecture & WECHAT & YOUKU & DNS & HTTP & TLS \\
\hline
 Skylake & $12.9\mu s$ & $15.0\mu s$ & $1.3\mu s$ & $3.3\mu s$ & $2.5\mu s$\\
\hline
Icelake & $10.7\mu s$ & $12.2\mu s$ & $0.9\mu s$ & $2.6\mu s$ & $2.0\mu s$\\
\hline
 Reduction & $17\%$ & $19\%$ & $31\%$ & $21\%$ & $20\%$\\
\hline
\end{tabular}
\label{tab:etala}
\end{table}

We also test the throughput with YOUKU. The average packets per flow is $20$ and more than $99\%$ flows are HTTP and TLS flows. The throughput is $6.5$\textit{Gbps} ($1,629$\textit{kpps}) per core. Since the average packets per flow is $28$ in Internet~\cite{kim2004flow}, we can estimate our throughput can achieve $9.1$\textit{Gbps} in most cases. Moreover, the throughput of feature extraction can achieve $35.3$\textit{Gbps}.

\subsubsection{Throughput in 5G UPF}
\begin{figure}[htp]
\centering
\subfigure[QQ, MEITUAN, QQNEWS]{\includegraphics[width=0.48\textwidth]{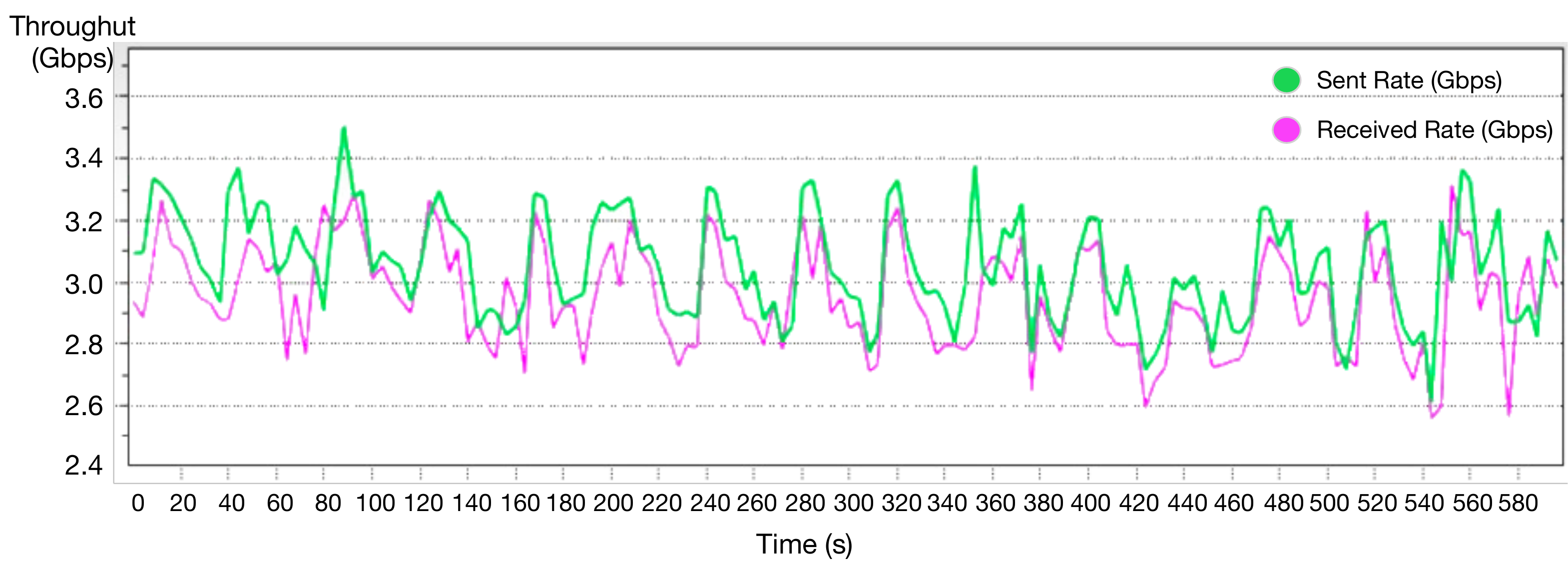}}
\hfill
\subfigure[QQ, TOUTIAO, HUOSHAN, KUAISHOU, QQNEWS]{\includegraphics[width=0.48\textwidth]{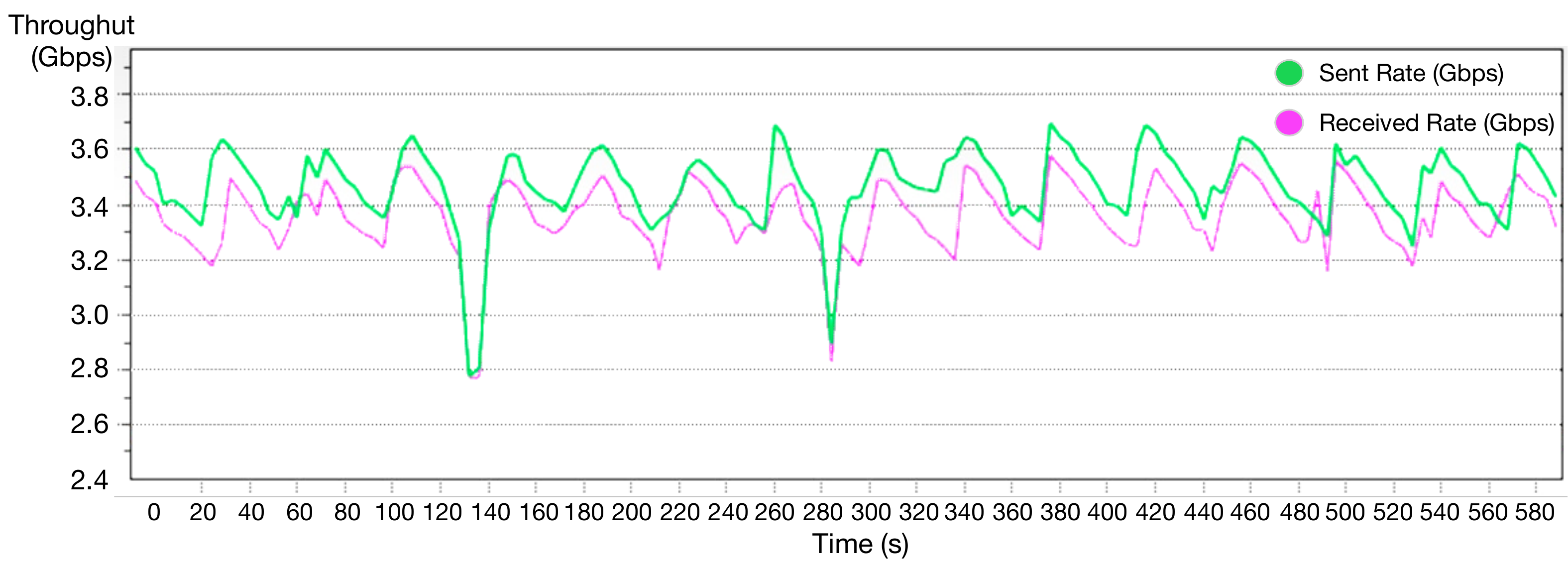}}
\hfill
\subfigure[BAIDU, TMALL, BILIBILI, TENCENT, TOUTIAO, KUAISHOU, QQ, HUOSHAN, QQNEWS]{\includegraphics[width=0.48\textwidth]{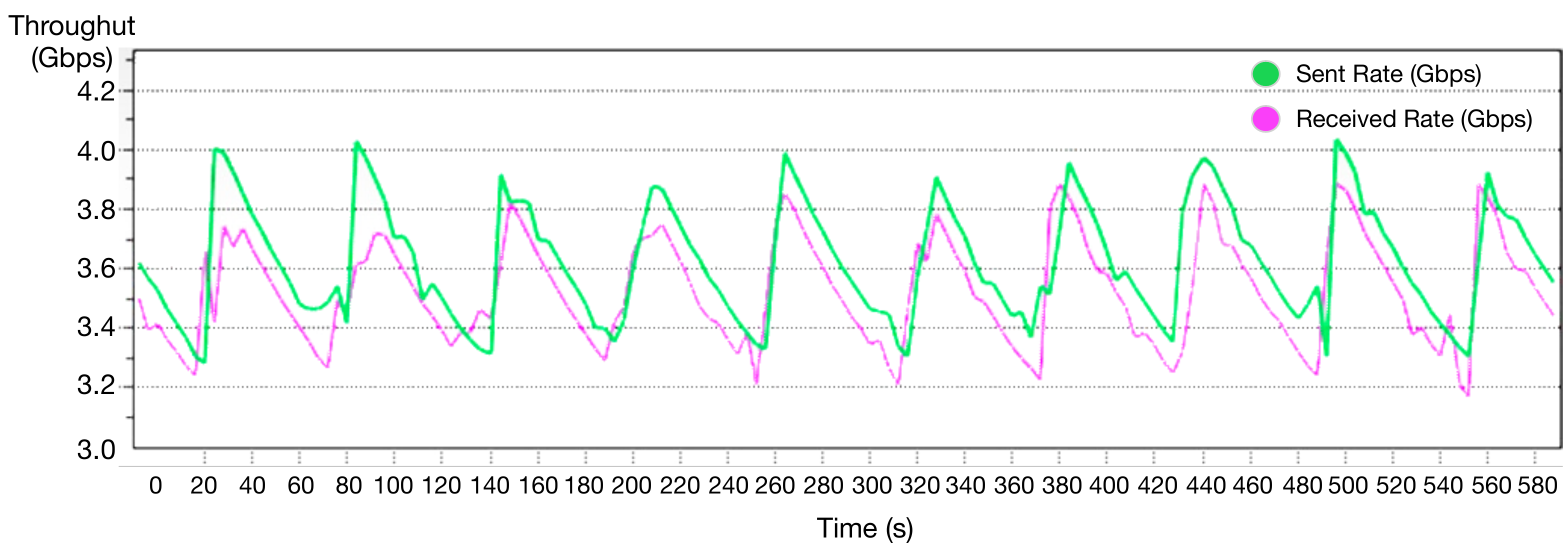}}
\caption{Throughput Evaluation with IXIA}
\label{fig:tp}
\end{figure}

We test our throughput with models that can classify $3$, $5$, and $9$ applications in 5G UPF respectively in Fig.~\ref{fig:tp}. The maximum throughput is $3.78$\textit{Gbps} ($618$\textit{kpps}) with $5$ applications and can get $3.39$\textit{Gbps} ($515$\textit{kpps}) and $3.58$\textit{Gbps} ($599$\textit{kpps}) with $3$ and $9$ applications. The result shows that the performance will not reduce with the increasing number of applications. The reason why the throughput in 5G UPF cannot achieve our aforementioned throughput and it has jitter is our nai\"ve flow table implementation and other integration overhead.

\subsection{Malware Detection}
\subsubsection{Accuracy}
We implement a ModSecurity plugin for SQLi/XSS with TADK. We compare our plugin with the well-utilized \textit{libinjection}~\cite{libin} in the same server environment (Nginx with ModSecurity). We set an attacking client with \textsc{SQLMAP} and \textsc{XSStrike} to generate traffic to test the accuracy. TADK's plugin has higher accuracy ($100\%$ for SQLi and $99.8\%$ for XSS) than the libinjection and it has fewer false positives.

\subsubsection{Latency}
We evaluate the latency of SQLi/XSS plugin in Table~\ref{tab:malla}. 
TADK's latency is $50\%$ less than libinjection.
In conclusion, The AI-based solution has lower latency than a rule-based solution in SQLi/XSS that makes real-time AI-based malware detection possible.

\begin{table}[!htp]\footnotesize
\caption{Latency per request}
\centering
\begin{tabular}{|c|c|c|c|c|c|}
\hline
Plugin & SQL injection & Cross-Site Script \\
\hline
 libinjection & $14.4\mu s$ & $8.9\mu s$\\
\hline
 TADK & $6.1\mu s$ & $4.5\mu s$\\
\hline
\end{tabular}
\label{tab:malla}
\end{table}

\section{Conclusion}
In this paper, we proposed TADK as a solution to address real-time AI-based networking workloads processing. The evaluation result shows that the application implemented with TADK can meet the requirements for real-time performance ($4.5\mu s$ per request on malware detection, $6.5$\textit{Gbps} and $35.3$\textit{Gbps} per core on traffic classification and feature extraction), accuracy ($\ge 95\%$), and scalability without any specialized hardware. We have deployed our solution in WAF and 5G UPF and we have evaluated it for real-world usage. We are currently working with our partners to improve the reliability and missing features (e.g., GQUIC) required for real-world deployment, and will be examined and used by the public and community eventually.
\label{sec:con}

\bibliographystyle{IEEEtran}
\bibliography{main}

\end{document}